\documentstyle[12pt]{article}

\def\hybrid{\topmargin -20pt    \oddsidemargin 0pt
        \headheight 0pt \headsep 0pt
        \textwidth 6.25in       
        \textheight 9.5in       
        \marginparwidth .875in
        \parskip 5pt plus 1pt   \jot = 1.5ex}

\hybrid

\def\baselinestretch{1.2}

\catcode`\@=11

\def\marginnote#1{}
\newcount\hour
\newcount\minute
\newtoks\amorpm
\hour=\time\divide\hour by60 \minute=\time{\multiply\hour by60
\global\advance\minute by-\hour}
\edef\standardtime{{\ifnum\hour<12 \global\amorpm={am}%
        \else\global\amorpm={pm}\advance\hour by-12 \fi
        \ifnum\hour=0 \hour=12 \fi
        \number\hour:\ifnum\minute<10 0\fi\number\minute\the\amorpm}}
\edef\militarytime{\number\hour:\ifnum\minute<10
0\fi\number\minute}

\def\draftlabel#1{{\@bsphack\if@filesw {\let\thepage\relax
   \xdef\@gtempa{\write\@auxout{\string
      \newlabel{#1}{{\@currentlabel}{\thepage}}}}}\@gtempa
   \if@nobreak \ifvmode\nobreak\fi\fi\fi\@esphack}
        \gdef\@eqnlabel{#1}}
\def\@eqnlabel{}
\def\@vacuum{}
\def\draftmarginnote#1{\marginpar{\raggedright\scriptsize\tt#1}}

\def\draft{\oddsidemargin -.5truein
        \def\@oddfoot{\sl preliminary draft \hfil
        \rm\thepage\hfil\sl\today\quad\militarytime}
        \let\@evenfoot\@oddfoot \overfullrule 3pt
        \let\label=\draftlabel
        \let\marginnote=\draftmarginnote
   \def\@eqnnum{(\theequation)\rlap{\kern\marginparsep\tt\@eqnlabel}%
\global\let\@eqnlabel\@vacuum}  }

\def\preprint{\twocolumn\sloppy\flushbottom\parindent 2em
        \leftmargini 2em\leftmarginv .5em\leftmarginvi .5em
        \oddsidemargin -.5in    \evensidemargin -.5in
        \columnsep .4in \footheight 0pt
        \textwidth 10.in        \topmargin  -.4in
        \headheight 12pt \topskip .4in
        \textheight 6.9in \footskip 0pt
        \def\@oddhead{\thepage\hfil\addtocounter{page}{1}\thepage}
        \let\@evenhead\@oddhead \def\@oddfoot{} \def\@evenfoot{} }

\def\numberbysection{\@addtoreset{equation}{section}
        \def\theequation{\thesection.\arabic{equation}}}

\def\underline#1{\relax\ifmmode\@@underline#1\else
        $\@@underline{\hbox{#1}}$\relax\fi}

\def\titlepage{\@restonecolfalse\if@twocolumn\@restonecoltrue\onecolumn
     \else \newpage \fi \thispagestyle{empty}\c@page\z@
        \def\thefootnote{\fnsymbol{footnote}} }

\def\endtitlepage{\if@restonecol\twocolumn \else \newpage \fi
        \def\thefootnote{\arabic{footnote}}
        \setcounter{footnote}{0}}  

\catcode`@=12 \relax

\def\figcap{\section*{Figure Captions\markboth
        {FIGURECAPTIONS}{FIGURECAPTIONS}}\list
        {Figure \arabic{enumi}:\hfill}{\settowidth\labelwidth{Figure
999:}
        \leftmargin\labelwidth
        \advance\leftmargin\labelsep\usecounter{enumi}}}
 \relax
\def\tablecap{\section*{Table Captions\markboth
        {TABLECAPTIONS}{TABLECAPTIONS}}\list
        {Table \arabic{enumi}:\hfill}{\settowidth\labelwidth{Table
999:}
        \leftmargin\labelwidth
        \advance\leftmargin\labelsep\usecounter{enumi}}}
 \relax
\def\reflist{\section*{References\markboth
        {REFLIST}{REFLIST}}\list
        {[\arabic{enumi}]\hfill}{\settowidth\labelwidth{[999]}
        \leftmargin\labelwidth
        \advance\leftmargin\labelsep\usecounter{enumi}}}
 \relax
\makeatletter
\newcounter{pubctr}
\def\publist{\@ifnextchar[{\@publist}{\@@publist}}
\def\@publist[#1]{\list
        {[\arabic{pubctr}]\hfill}{\settowidth\labelwidth{[999]}
        \leftmargin\labelwidth
        \advance\leftmargin\labelsep
        \@nmbrlisttrue\def\@listctr{pubctr}
        \setcounter{pubctr}{#1}\addtocounter{pubctr}{-1}}}
\def\@@publist{\list
        {[\arabic{pubctr}]\hfill}{\settowidth\labelwidth{[999]}
        \leftmargin\labelwidth
        \advance\leftmargin\labelsep
        \@nmbrlisttrue\def\@listctr{pubctr}}}
 \relax
\makeatother
\newskip\humongous \humongous=0pt plus 1000pt minus 1000pt

\newif\ifdtup

\relax

\def\be{\begin{equation}}
\def\ee{\end{equation}}
\def\ba{\begin{eqnarray}}
\def\ea{\end{eqnarray}}

\begin{document}

\renewcommand{\theequation}{\arabic{equation}}

\newcommand{\beq}{\begin{equation}}
\newcommand{\eeq}[1]{\label{#1}\end{equation}}
\newcommand{\ber}{\begin{eqnarray}}
\newcommand{\eer}[1]{\label{#1}\end{eqnarray}}
\newcommand{\eqn}[1]{(\ref{#1})}
\begin{titlepage}
\begin{center}

\hfill hep--th/0501127\\
\vskip -.1 cm
\hfill CERN-PH-TH/2005-004\\
\vskip -.1 cm
\hfill UO-HET/515\\
\vskip -.1 cm
\hfill January 2005\\

\vskip .4in

{\large \bf Aspects of WZW models at critical
level}\footnote{Based on talks presented by I.B. at the {\em 37th
International Symposium Ahrenshoop on the Theory of Elementary
Particles}, Berlin, August 23-27, 2004 and by C.S. at the {\em RTN
and EXT Meeting on the Structure of Spacetime}, Kolymbari,
September 5-10, 2004; to appear in the proceedings published in
Fortschritte der Physik.}

\vskip 0.3in

{\bf Ioannis Bakas}\footnote{On sabbatical leave from Department
of Physics, University of Patras, GR-26500 Patras, Greece; e-mail:
bakas@ajax.physics.upatras.gr} \vskip 0.1in
{\em Theory Division, Department of Physics, CERN\\
CH-1211 Geneva 23, Switzerland\\
\footnotesize{\tt ioannis.bakas@cern.ch}}\\

\vskip 0.3in

{\bf Christos Sourdis} \vskip 0.1in
{\em Department of Physics, Graduate School of Science, Osaka University\\
Toyonaka, Osaka 560-0043, Japan\\
\footnotesize{\tt sourdis@het.phys.sci.osaka-u.ac.jp}}\\

\end{center}

\vskip .6in

\centerline{\bf Abstract}

\noindent We consider non-compact WZW models at critical level
(equal to the dual Coxeter number) as tensionless limits of
gravitational backgrounds in string theory. Special emphasis is
placed on the Euclidean black hole coset $SL(2,R)_k/U(1)$ when
$k=2$. In this limit gravity decouples in the form of a Liouville
field with infinite background charge and the world-sheet symmetry
of the model becomes a truncated version of $W_{\infty}$ without
Virasoro generator. This is regarded as manifestation of Langlands
duality for the $SL(2,R)_k$ current algebra that relates small
with large values of the level in the two extreme limits. However,
the physical interpretation of the $SL(2,R)_k/U(1)$ coset model
below the self-dual value $k=3$ remains elusive including the
non-conformal theory at $k=2$. \vfill
\end{titlepage}
\eject

\def\baselinestretch{1.2}
\baselineskip 16 pt \noindent

Wess-Zumino-Witten (WZW) models based on non-compact groups $G_k$,
and their gauged versions, provide a wide class of exact conformal
field theory backgrounds (or building blocks thereof) for studying
string propagation in non-trivial space-time geometries. The level
$k$ of the underlying current algebra is typically constrained to
take continuous values $k > g^{\vee}$ by unitarity, where
$g^{\vee}$ is the dual Coxeter number given by the value of the
quadratic Casimir operator of $G$ in the adjoint representation.
The large $k$ limit dictates the semi-classical geometry of these
models by associating a metric, an antisymmetric tensor field and
a dilaton that satisfy the vanishing condition of the beta
function equations to lowest order in the perturbative
$\alpha^{\prime}$ expansion (see, for instance, \cite{ed}). Of
course, there can be subleading corrections to higher orders in
perturbation theory that are computable in closed form by
Hamiltonian methods (see, for instance, \cite{dvv}). They are in
accord with the perturbative renormalization of general non-linear
sigma models in two dimensions. In this group theoretical setting
the parameter $\alpha^{\prime}$ is supplied by $(k-g^{\vee})^{-1}$
and therefore WZW models at critical level $k=g^{\vee}$ can be
regarded as exact tensionless theories, \cite{ulf, sourd}. This
limit is ultra-quantum in nature and it is only possible to
consider for non-compact groups. Furthermore, it is a singular
limit within the class of conformal field theories, since the
central charge of the corresponding Virasoro algebra blows up to
infinity, but it makes perfect sense in the more general class of
two-dimensional quantum field theories.

We are set to examine the nature of WZW models at critical level
with emphasis on the simplest case of $SL(2,R)_k$ current algebra
for which $g^{\vee} = 2$. After a brief overview of the main
problems and shortcomings of the effective field theory of such
limiting models we will resort to world-sheet methods, which are
well defined for all values of the level, and use them to explore
their symmetries. Higher rank generalizations are also possible to
consider using the appropriate technical ingredients. For further
details on these and other related issues we refer the reader to
our published work, \cite{sourd}.

The simplest (yet interesting) example of this kind is provided by
the gauged WZW coset $SL(2,R)_k/U(1)$, which admits the
interpretation of a two-dimensional Euclidean black hole in the
large $k$ limit, \cite{ed}, \be ds^2 \simeq k \left(dr^2 + {\rm
tanh}^2 r ~ d\theta^2 \right) , ~~~~~ \Phi \simeq 2 {\rm log}
\left({\rm cosh} r \right), \ee since the semi-classical geometry
looks like a semi-infinite long cigar satisfying the equation
$R_{\mu \nu} = \nabla_{\mu} \nabla_{\nu} \Phi$ imposed by
conformal invariance to lowest order in perturbation theory.
Higher order corrections transform the background to the
perturbatively exact form, \cite{dvv}, \be ds^2 \simeq (k-2)
\left(dr^2 + \beta^2 (r) ~ d\theta^2 \right) , ~~~~~ \Phi \simeq
{\rm log} \left({{\rm sinh} 2r \over \beta(r)} \right) \ee for all
values of $k$, where $\beta(r)$ is given by \be \beta^{-2}(r) =
{\rm coth}^2 r - {2 \over k} ~. \ee Thus, by pretending that the
result is valid all the way down to the critical level, it readily
follows as $k \rightarrow 2$ that \be ds^2 \simeq (k-2) \left(dr^2
+ {\rm sinh}^2 r ~ d\theta^2 \right) , ~~~~~ \Phi \simeq {\rm log}
\left({\rm cosh} r \right) , \ee which corresponds to the geometry
of an infinitely curved hyperboloid in the ultra-quantum regime.

The reasoning of the effective theory should be taken with care
when the level of the current algebra becomes critical. It is well
known that in all tensionless theories there is an abundance of
states that become massless and as such they should be taken on
equal footing with the massless modes provided by the metric,
dilaton and anti-symmetric tensor field (where it is applicable).
For, it is not consistent to truncate the action only to the
generic massless modes without accounting for the contribution of
the ``will be massless" states in the ultra-quantum regime.
Otherwise, the effective geometry appears to be singular, as in
the case noted above. This is a well known phenomenon which cannot
be controlled here by the inclusion of a few extra modes, as in
other more tractable cases. As a result, there is no systematic
method to study WZW models at critical level, such as
$SL(2,R)_2/U(1)$, within the framework of the local effective
field theory, which is only suitable for summarizing the results
of the perturbative expansion. This shortcome is also consistent
with the observation that non-linear sigma models with non-compact
target spaces are non-perturbatively non-renormalizable in the
sense that there can be divergencies giving rise to counterterms
that contain higher dimensional operators and which correspond to
coupling of gravitons to massive fields, \cite{das}. These
divergencies arise when the $({\rm momenta})^2$ are equal or
larger than $1/\alpha^{\prime}$, in which case any truncation of
the perturbative expansion does not make sense. Hence, their
effect becomes dominant in the tensionless limit $\alpha^{\prime}
\rightarrow \infty$. All these provide indications that
non-commutative geometry might be a better framework to formulate
tensionless models by means of infinitely non-commutative
structures, but we will not pursue this line of thought further in
the present work.

There is an alternative description of the bosonic $SL(2,
R)_k/U(1)$ model for small values of $k$ (close to its critical
value) in terms of the sine-Liouville conformal field theory,
which is roughly \be S= {1 \over 4\pi} \int \sqrt{h} d^2x \left({1
\over k-2} (\partial r)^2 + {1 \over k} (\partial \theta)^2 - {1
\over k-2} R[h] r + \mu^2 e^{-r} {\rm cos} \theta \right), \ee
where $h$ is the world-sheet metric with Ricci scalar $R[h]$ and
$\mu$ some parameter, following a conjecture by Fateev,
Zamolodchikov and Zamolodchikov (FZZ) (for an overview see, for
instance, \cite{kut}). The relation between $SL(2,R)_k/U(1)$ and
the sine-Liouville models is a strong-weak coupling duality on the
world-sheet. The coset conformal field theory becomes weakly
coupled in the limit $k \rightarrow \infty$, where the cigar
geometry is weakly curved. On the other hand, the semi-classical
limit of the sine-Liouville theory arises as $k \rightarrow 2$,
where it becomes weakly coupled, unlike the coset space theory
that becomes strongly coupled. Both theories exhibit an asymptotic
region $r \rightarrow \infty$ where the geometry is that of a
cylinder with respective radii $\sqrt{k}$ and $1/\sqrt{k}$ with
linear dilaton. The potential term of the sine-Liouville action is
negligible for large $r$, as it decays exponentially, and the two
asymptotic geometries are simply related to each other by
$T$-duality along the angular direction. The non-trivial aspect of
the FZZ duality comes from the diminishing size of the $S^1$ fiber
as one approaches the tip of the cigar, which in turn modifies the
naive duality rules and it accounts for the generation of the
potential term by vortex-instanton effects on the sine-Liouville
side using mirror symmetry. Of course, such relation has only been
proven rigorously for a supersymmetric variant of these models,
namely for the corresponding Kazama-Suzuki coset and $N=2$
Liouville theory, \cite{hori}, as originally proposed in
\cite{G-K}. Nevertheless, since the potential term is unbounded
from below as $r$ ranges on the whole real axis, it is natural to
expect that for small values of $k$ the semi-classical reasoning
can be employed to render the sine-Liouville model unstable. This
observation is also consistent with the fact that the coset theory
is well defined only for $k>2$.

The limiting value $k=2$ appears to be rather special for both
sides of the FZZ correspondence as the central charge of the
Virasoro algebra \be c=2 ~ {k+1 \over k-2} \ee becomes infinite.
It manifests in the form of a Liouville field with infinite
background charge that effectively decouples from the rest of the
spectrum leading to a contraction of the Virasoro algebra to an
abelian structure. The precise nature of the remnant model is not
completely understood to this date but some of its symmetry
aspects can be revealed using world-sheet methods. Its proper
interpretation in the framework of quantum field theories will
also help to explore the gravitational undressing of more general
string backgrounds, for which an equivalent description as
Liouville theory coupled to other matter fields is only valid
asymptotically. Note that the contraction of the Virasoro algebra
is also common to other tensionless string models and as a result
there is no need to have critical space-time dimensionality for
consistent string propagation (see, for instance, \cite{ulf} and
references therein). The meaning of space-time geometry itself is
also questionable in the tensionless limit, but there is no better
definition of string theory at this moment which can accommodate
the notion of tensionless objects in a background independent way.

Ideally, we would like to have a field theoretic manifestation of
Langlands (electric-magnetic) duality for $SL(2,R)_k$ current
algebra with a level relation (see, for instance, \cite{ff})
\begin{equation}
k^{\prime} - 2 = {1 \over k-2} ~. \label{lang}
\end{equation}
Thus, the limit that $k$ takes its critical value on one side
could be compared directly with the large $k$ limit on another
side, where the space-time interpretation of the theory is valid.
In both cases there are important simplifications that occur and
certainly the FZZ correspondence is pointing to the right
direction by comparing two different models. However, a drawback
of the current formalism is the apparent instability of the
sine-Liouville theory for small values of $k$ that might also
include $k=2$. Also, it is not clear what is the phase of the
remnant theory after decoupling of the Liouville field which
carries infinite background charge. One might naively think that
such decoupling will remove the infinity from the Virasoro central
charge and leave the conformal symmetry intact but with finite
(possibly zero) central charge. It has been suggested in a
different context that $SL(2,R)_2/U(1)$ might be a topological
conformal field theory with zero net central charge that arises by
taking an average value prescription for $c$ as $k \rightarrow
2^{\pm}$, \cite{john}. The subsequent analysis of the world-sheet
symmetries of the coset model shows that there is no such remnant
of the conformal symmetry and thus its interpretation remains an
open question.

We further note here that according to the relation \eqn{lang}
there is a {\em self-dual} value of the level for
\begin{equation}
k = k^{\prime} = 3 ~.
\end{equation}
This value provides the boarder line for the existence of
normalizable zero modes of the $SL(2,R)_k/U(1)$ coset associated
to marginal operators with dimension 1 (see, for instance,
\cite{hori} for a complete discussion). For $k>3$ the existence of
this zero mode prohibits changes of the asymptotic radius of the
cigar geometry (given by $\sqrt{k}$ in appropriate units), whereas
for $k<3$ its absence seems to allow for such possibility.
Clearly, this is a very important difference that holds the key
for the physical interpretation of the $SL(2,R)_k/U(1)$ coset for
small values of $k$ (including $k=9/4$ with Virasoro cental charge
$c = 26$), \cite{malda}; it also shows up in recent attempts to
understand the black hole (non)-formation in the singlet sector of
matrix models, \cite{andy}. Finally note that the $N=2$
supersymmetric case appears to be more tractable in many respects
with the relation \eqn{lang} being replaced by $k \rightarrow
1/k$. Then, the critical level is $k=0$, while the self-dual value
is shifted to $k=1$. It will also be interesting (and possibly
easier) to understand the field theoretic description of the
supersymmetric coset at $k=0$ given its equivalence to $N=2$
Liouville theory.

We think that better understanding of the Langlands duality for
current algebras and its implications for gauged WZW models will
help to clarify the situation. In this respect, the ultimate
understanding of $SL(2,R)_k/U(1)$ at critical level seems to be
lying at the heart of the problem. In the following we restrict
attention to the world-sheet symmetries of the coset model
$SL(2,R)_k/U(1)$ and examine the structure of its $W$-algebra in
two extreme cases $k=\infty$ and $k=2$ that correspond to the
semi-classical gravitational limit and the tensionless limit,
respectively. It will be shown with the aid of non-compact
parafermions, and their operator product expansion, that the
world-sheet symmetry linearizes in both limits and can be
subsequently identified with the $W_{\infty}$ algebra and its
non-conformal higher spin truncation, respectively, \cite{sourd}.
This result can be considered as a special manifestation of
Langlands duality, but it is not clear how it generalizes for
intermediate values of $k$. The parafermion currents do not seem
to have a definite transformation rule among themselves under
\eqn{lang}, which in turn prevents to relate the structure of the
corresponding $W$-algebras for general values of $k$. After all,
given the FZZ correspondence (or mirror symmetry in the $N=2$
supersymmetric case), the field theoretic realization of such
duality is not manifest within the same coset, but it rather
relates different classes of models.

Recall that the lowest parafermions currents of the non-compact
WZW model, $\psi_{\pm 1}(z)$, admit the following realization in
terms of two-dimensional free fields, \cite{lyk, peskin, bak},
\begin{equation}
\psi_{\pm 1}(z) = {1 \over \sqrt{2k}} \left(\mp \sqrt{k-2}
\partial \phi_1 (z) + i \sqrt{k} \partial \phi_2 (z) \right) {\rm
exp} \left(\pm i \sqrt{{2 \over k}} \phi_2 (z) \right)
\end{equation}
for all $k \geq 2$. The fields $\phi_i(z)$ are both space-like
with normalized two-point functions
\begin{equation}
<\phi_i (z) \phi_j (w)> = - \delta_{ij} {\rm log}(z-w) ~.
\end{equation}
The operator product expansion of parafermions assumes the form
\begin{equation}
\psi_{+1}(z) \psi_{-1}(w) = {1 \over (z-w)^{2\Delta}} \left(1 + {2
\Delta \over c_{\psi}} (z-w)^2 T_{\psi}(w) + {\cal O}(z-w)^3
\right) , \label{ope}
\end{equation}
where $\Delta$ is the conformal dimension of $\psi_{\pm 1}(z)$ and
$c_{\psi}$ is the central charge of the Virasoro algebra with
values
\begin{equation}
\Delta = 1 + {1 \over k} ~, ~~~~~ c_{\psi} = 2{k+1 \over k-2} ~.
\end{equation}
The associated stress-energy tensor of the parafermion theory is
represented as
\begin{equation}
T_{\psi} (z) = -{1 \over 2} (\partial \phi_1)^2 - {1 \over 2}
(\partial \phi_2)^2 + {1 \over \sqrt{2(k-2)}} \partial^2 \phi_1 ~,
\end{equation}
whereas the higher order terms in the expansion \eqn{ope} give
rise to higher conserved chiral currents and their free field
realization in terms of $\phi_1(z)$ and $\phi_2(z)$. All these
currents generate the extended world-sheet symmetry of the
parafermion theory, which is also known as $W$-algebra.

The parafermions of the coset can be dressed with an additional
$U(1)$ field $\chi(z)$, which is assumed to have two-point
function $<\chi(z) \chi(w)> = - {\rm log}(z-w)$, and the composite
fields
\begin{equation}
J^{\pm}(z) = \sqrt{k} \psi_{\pm 1}(z) {\rm exp} \left(\pm \sqrt{{2
\over k}} \chi(z) \right) , ~~~~ J^3 (z) = -\sqrt{{k \over 2}}
\partial \chi(z)
\end{equation}
satisfy the $SL(2,R)_k$ current algebra with level $k \geq 2$,
\cite{peskin}, since
\begin{eqnarray}
J^+(z) J^-(w) & = & {k \over (z-w)^2} -2{J^3(w) \over z-w} ~, \nonumber\\
J^3(z) J^{\pm}(w) & = & \pm {J^{\pm}(w) \over z-w} ~, \nonumber\\
J^3(z) J^3(w) & = & -{k \over 2(z-w)^2} ~,
\end{eqnarray}
up to non-singular terms. In this case, the Sugawara construction
for the $SL(2,R)_k$ current algebra yields the stress-energy
tensor
\begin{equation}
T(z) = {1 \over 2(k-2)} \left(J^+ J^- + J^- J^+ - 2 J^3 J^3
\right)(z) = -{1 \over 2}(\partial \chi)^2 + T_{\psi}(z)
\end{equation}
with total Virasoro central charge $c= 1+c_{\psi}$. At $k=2$ the
center of the enveloping of the current algebra becomes
non-trivial, as it is generated by $(J^+ J^- + J^- J^+ - 2 J^3 J^3
)(z)$, and it accounts for the contraction of the Virasoro algebra
to an abelian structure.

Next, let us examine the parafermion currents in the large $k$
limit. They assume the following simple form,
\begin{equation}
\psi_{+1}(z) = -{1 \over \sqrt{2}} \partial \bar{\Phi} (z) ~,
~~~~~ \psi_{-1}(z) = {1 \over \sqrt{2}} \partial \Phi (z) ~,
\label{bf1}
\end{equation}
where $\Phi(z)$ is a complex free boson taken to be $\Phi = \phi_1
+ i \phi_2$. In this case, the parafermions become local abelian
currents of dimension $\Delta =1$ and their operator product
expansion \eqn{ope} can only give rise to boson bilinear terms of
the form $\partial^k \bar{\Phi} \partial^{s-k} \Phi (z)$ to each
order $s$ in the $z-w$ expansion with $k = 1, \cdots , s-1$. A
convenient quasi-primary basis of the $W$-algebra generators is
chosen to be \cite{sourd, bak}
\begin{equation}
W_s(z) = 2^{s-4} s{(s-2)! \over (2s-3)!!} \sum_{k=1}^{s-1} (-1)^k
{s-1 \choose k} {s-1 \choose k-1} \partial^k \bar{\Phi}
\partial^{s-k} \Phi(z) \label{form1}
\end{equation}
describing fields for all integer values of spin $s\geq 2$. For
$s=2$, in particular, one recovers the stress-energy tensor
$T_{\psi}(z) = W_2(z)$ of the coset model in the large $k$ limit.
The operators $W_s(z)$ generate a linear infinite dimensional
algebra which is identified with $W_{\infty}$.

On the other hand, at the critical level $k=2$, there is a drastic
reduction in the structure of parafermion currents as one of the
two free bosons decouples completely. In this case the
parafermions assume the form
\begin{equation}
\psi_{+1}(z) = {1 \over \sqrt{2}} \partial \bar{\Psi} (z) ~, ~~~~~
\psi_{-1}(z) = - {1 \over \sqrt{2}} \partial \Psi (z) ~,
\label{bf2}
\end{equation}
where
\begin{equation}
\Psi(z) = e^{-i\phi_2(z)} ~, ~~~~~ \bar{\Psi}(z) = e^{i\phi_2(z)}
\end{equation}
are fermionic currents that are described as vertex operators of
the remaining free boson. Then, since $<\bar{\Psi}(z) \Psi(w)> =
1/(z-w)$, it follows that the operator product expansion \eqn{ope}
gives rise to fermion bilinear terms of the form $\partial^k
\bar{\Psi} \partial^{s-k-1} \Psi(z)$ to each order $s$ in the
$z-w$ expansion with $k=1, \cdots, s-2$. However, unlike the
previous (semi-classical) case, there is no contribution to order
$s=2$ because the coefficient $2\Delta / c_{\psi} \rightarrow 0$
as $k \rightarrow 2$. This is also consistent with the absence of
conformal symmetry for WZW models at critical level. As for the
remaining generators which arise to higher orders in the
expansion, it is convenient to choose the basis, \cite{sourd},
\begin{equation}
\tilde{W}_s(z) = 2^{s-3} s(s+1){(s-3)! \over (2s-3)!!}
\sum_{k=1}^{s-2} (-1)^{k+1} {s-1 \choose k+1} {s-1 \choose k-1}
\partial^{k} \bar{\Psi} \partial^{s-k-1} \Psi (z) \label{form2}
\end{equation}
for all $s \geq 3$ and identify their algebra with a consistent
truncation of $W_{\infty}$ by moding out the Virasoro algebra.

The above identifications can be made precise by considering the
class of infinite dimensional Lie algebras of $W_{\infty}$ type,
\cite{prs},
\begin{eqnarray}
[V_n^s , ~ V_m^{s^{\prime}}] & = & \left((s^{\prime}-1) n - (s-1)m
\right) V_{n+m}^{s+s^{\prime}-2} + \sum_{r\geq 1}
g_{2r}^{ss^{\prime}}(n,m; \mu)
V_{n+m}^{s+s^{\prime}-2-2r} + \nonumber\\
& & c_s(\mu) n(n^2-1)(n^2-4) \cdots (n^2-(s-1)^2)
\delta_{s,s^{\prime}} \delta_{n+m, 0} ~.
\end{eqnarray}
The commutation relations of the Fourier modes can be cast into
operator product expansions for the corresponding currents
\begin{equation}
V_n^s = \oint_0 {dz \over 2\pi i} z^{n+s-1} V_s(z)
\end{equation}
and employ contour integration as in two dimensional field
theories. In either case, the structure constants of the algebra
are determined by the following expressions,
\begin{equation}
g_{2r}^{ss^{\prime}}(n,m; \mu) = {\phi_{2r}^{ss^{\prime}}(\mu)
\over 2 (2r+1)!} N_{2r}^{ss^{\prime}}(n,m) ~,
\end{equation}
where
\begin{equation}
\phi_{2r}^{ss^{\prime}}(\mu) = \sum_{k=0}^{r} {(-{1 \over 2}
-2\mu)_k ({3 \over 2} + 2\mu)_k (-r-{1 \over 2})_k (-r)_k \over k!
(-s+ {3 \over 2})_k (-s^{\prime} + {3 \over 2})_k
(s+s^{\prime}-2r- {3 \over 2})_k}
\end{equation}
and
\begin{equation}
N_{2r}^{ss^{\prime}}(n,m) = \sum_{k=0}^{2r+1} (-1)^k {2r+1 \choose
k} (2s-2r-2)_k [2s^{\prime}-k-2]_{2r+1-k} [s-1+n]_{2r+1-k}
[s^{\prime}-1+m]_k ~.
\end{equation}
Finally, the ascending and descending Pochhammer symbols are
defined as usual,
\begin{equation}
(a)_n = a(a+1) \cdots (a+n-1) ~, ~~~~~ [a]_n = a(a-1) \cdots
(a-n+1) ~,
\end{equation}
whereas the coefficients of the central terms are given
collectively by the expression
\begin{equation}
c_s(\mu) = 2^{2(s+\mid \mu \mid) -7} c {(s+2\mu)! (s-2\mu -2)!
\over (2s-1)!! (2s-3)!!} ~.
\end{equation}

The class of these algebras depends on a parameter $\mu$ that can
take all values $\mu = -1/2, 0, 1/2, 1, \cdots$. For each such
value the operator content of the algebra truncates consistently
to all integer spin $s \geq 2\mu + 2$, since the structure
constants vanish otherwise. Thus, for $\mu = -1/2$ the algebra is
$W_{1+\infty}$, for $\mu = 0$ it is $W_{\infty}$, whereas for
higher values of $\mu$ the resulting algebras do not contain the
Virasoro generators nor any other fields with spin less than $2\mu
+2$ in their spectrum. Thus, in this context we find that the
$W$-algebra of the $SL(2,R)_k/U(1)$ coset model at $k=\infty$
corresponds to $\mu = 0$ with $V_s(z) = W_s(z)$ as given by
equation \eqn{form1} and the (Virasoro) central charge is $c=2$
equal to the dimension of the classical black hole geometry.
Likewise, the $W$-algebra of the coset model at $k=2$ corresponds
to $\mu = 1/2$ with $V_s(z) = \tilde{W}_s(z)$ as given by equation
\eqn{form2} and the coefficient of the central terms turns out to
be $c=2$ using the normalization given above, \cite{sourd}. In
this case, however, the value of the central charge does not have
a direct physical interpretation in the absence of Virasoro
generators.

The resulting algebraic structures are very similar to each other
following by consistent truncation of the Virasoro generators.
Actually, the two limiting cases are formally related at the level
of parafermion currents $\psi_{\pm 1}$ by the bose-fermi relation
(up to a sign)
\begin{equation}
\Phi(z) \leftrightarrow \Psi(z) ~,
\end{equation}
as can be readily seen from equations \eqn{bf1} and \eqn{bf2}
above. We view this result as manifestation of Langlands duality
with the level relation \eqn{lang} in the two extreme limits of
its validity. Then, the form of the corresponding $W$-generators
\eqn{form1} and \eqn{form2} simply follows by performing the
operator product expansions and choosing appropriate basis to
diagonalize the central terms of the algebras. It will be
interesting to understand the meaning of this result in a wider
context and extend its description beyond the two limiting values
$k=2$ and $k=\infty$ in an appropriate way.

The decoupling of the Liouville field at critical level is one of
the most important aspects of WZW models for non-compact groups.
Essentially, it rips off one space-time dimension and takes the
remnant theory in a non-conformal phase. This disintegration is
also seen using other free field realizations of the $SL(2,R)_k$
current algebra, such as the standard Wakimoto representation,
\cite{waki},
\begin{eqnarray}
& & J^- (z) = \beta(z) ~, ~~~~ J^3(z) = \beta \gamma + \sqrt{{k-2
\over 2}}
\partial \varphi (z) ~ , \nonumber\\
& & J^+(z) = \beta \gamma^2 (z) + \sqrt{2(k-2)} \gamma \partial
\varphi (z)+ k\partial \gamma (z) ~,
\end{eqnarray}
where $(\beta, \gamma)$ are commuting ghost fields with dimensions
$(1, 0)$ and $\varphi$ is a space-like boson, all having
normalized two-point functions. When $k>2$ three fields are
required to provide a faithful representation of the algebra,
whereas for $k=2$ only two suffice as the contribution of
$\varphi$ to the currents vanishes. The stress-energy tensor
following from the Sugawara construction is expressed as
\begin{equation}
T(z) = \beta \partial \gamma - {1 \over 2} (\partial \varphi)^2 -
{1 \over \sqrt{2(k-2)}} \partial^2 \varphi \label{wakirep}
\end{equation}
and therefore $\varphi$ is a Liouville field with infinite
background charge as $k \rightarrow 2$. However, after the
decoupling the theory is not the ordinary $(\beta, \gamma)$
conformal system. Thus, the WZW model and its gauged version are
both suffering a drastic change at critical level.

In view of the disappearance of all $\varphi$ dependence from the
bosonic realization \eqn{wakirep}, a complementary understanding
of this essential change may also be provided through the use of
restricted Wakimoto modules (see, for instance, \cite{boris} and
references therein) to investigate the singular structure of the
highest weight modules for $k=2$. For this limiting level value,
the Kac-Kazhdan equation degenerates to an identity and the
corresponding affine modules contain an infinite number of null
states obtained by the application of the commuting - at this
limit -  Virasoro modes, leading to a vanishing determinant
formula, \cite{kac} (but see also \cite{peskin}). This generic
feature of all affine Lie algebras at criticality is in full
accordance with the aforementioned decoupling of the Virasoro
generator from the spectrum of the $W$-algebra as the level
approaches its critical value.

There is another disintegration limit of the $SL(2,R)_k/U(1)$
coset model that works in the opposite direction and arises even
semi-classically. It is based on the observation that when the
$U(1)$ gauge group is boosted by an infinite amount the resulting
theory is equivalent to the null gauged WZW model $SL(2,R)_k/E(1)$
that describes the dynamics of the Liouville field alone. This
reduction of dimensionality can also be reached in a controlled
way by considering a very large boost, in which case the
semi-classical metric describes $c=1$ matter coupled to Liouville
field acting as a slowly varying background, \cite{iran}. In
either case, the theory remains conformal all the way up to its
disintegration limit for all $k>2$. At $k=2$ the null gauged WZW
model describes a Liouville field with infinite background charge
but there is no other remnant as in the ordinary case. It will be
interesting to examine the manifestation of the boosting in the
framework of the FZZ correspondence and in particular for the
$N=2$ mirror models for all values of $k$ including the critical
level.

The breakdown of the conformal invariance which is seen at
critical level might be similar in nature to the
Kosterlitz-Thouless transition of a $c=1$ free boson compactified
on a circle. In that case, the compactness of the target space
allows for the existence of vortices which are irrelevant at large
values of the radius, but they break conformal invariance at small
values; in fact the theory is equivalent to the sine-Gordon model.
It is conceivable that non-perturbative effects in the $SL(2,
R)_k/U(1)$ coset, as those considered in \cite{yung}, could play a
prominent role in this respect. Further support towards this
analogy is also provided by the free field realization of the
coset model at critical level in terms of a single compact space
boson. Starting from the semi-classical regime one may look at the
problem by considering the asymptotic radius of the Euclidean
black hole which is $\sqrt{k}$ and is shrinking as $k$ is lowered.
Although the semi-classical reasoning should not be further
trusted due to the perturbative $1/k$ corrections to the
background geometry, one is led to suspect that a transition might
occur to non-conformal phase for sufficiently small $k$. The
absence of normalizable marginal deformations that already occur
for $k \leq 3$ also shows that the physical interpretation of the
$SL(2,R)_k/U(1)$ coset should be modified for small values of $k$,
since changes of the radius become possible without affecting the
dilaton (unlike $k > 3$); actually, at $k=2$ all the discrete
representations are squeezed out from the spectrum, following
\cite{oogu}, thus signaling a transition caused by the liberation
of the bound states. Although the FZZ correspondence provides an
alternative way to look at this problem its resolution is still
lacking and calls for further work.

\newpage
\centerline{\bf Acknowledgments} \noindent This work was supported
in part by the European Research and Training Networks
``Superstring Theory" (HPRN-CT-2000-00122), ``Quantum Structure of
Space-time and the Geometric Nature of Fundamental Interactions"
(HPRN-CT-2000-00131) and ``Constituents, Fundamental Forces and
Symmetries of the Universe" (MRTN-CT-2004-005104). One of us
(C.S.) is also thankful to the Japanese Ministry of Education,
Culture, Sports, Science and Technology (Monbukagakusho) for
financial support during his stay in Osaka.

\end{document}